\documentclass[10pt]{article}

\usepackage{soul,color}
\usepackage{multicol}
\usepackage{graphicx}
\usepackage{amsmath,amssymb}
\usepackage[a4paper]{geometry}
\usepackage[colorlinks=true, linkcolor=blue, citecolor=blue, pdfborder={0 0 0}]{hyperref}
\usepackage{multirow}
\usepackage{makecell}
\usepackage{caption}
\usepackage[normalem]{ulem}

\newcommand\nustar{{\it NuSTAR}}

\newcommand\rxte{{\it RXTE}}

\newcommand\astrosat{{\it AstroSat}}
\newcommand\xmm{{\it XMM-Newton}}

\newcommand\nat{{\it Nature}}
\newcommand\apss{{\it Astrophysics and Space Science}}
\newcommand\apj{{\it The Astrophysical Journal}}
\newcommand\aap{{\it Astronomy \& Astrophysics}}
\newcommand\apjl{{\it The Astrophysical Journal Letter}}
\newcommand\mnras{{\it Monthly Notices of the Royal Astronomical Society}}
\newcommand\pasj{{\it Publications of the Astronomical Society of Japan}}

\begin{document}
%\onecolumn
\begin{center}
{\Large \bf Energy-dependent temporal study of GX 13+1 with \emph{AstroSat} observation.}

\vskip1.0cm

Arbind Pradhan$^{1,}${\footnote{E-mail:app22111@tezu.ac.in}},
Sree Bhattacherjee,$^{1,}${\footnote{E-mail: app21101@tezu.ac.in}},
Biplob Sarkar$^{1,}${\footnote{Corresponding author. E-mail:biplobs@tezu.ac.in}}
%fuhuliu@163.com; fuhuliu@sxu.edu.cn}}, Khusniddin K.
%Olimov$^{2,3,}${\footnote{Corresponding author. E-mail:
%khkolimov@gmail.com; kh.olimov@uzsci.net}}

{\small\it $^1$Department of Applied Sciences, Tezpur University, Tezpur, Assam, 784028}
	
%$^2$Laboratory of High Energy Physics, Physical-Technical
%Institute of Uzbekistan Academy of Sciences, \\ Chingiz Aytmatov
%Str. 2b, Tashkent 100084, Uzbekistan

%$^3$Department of Natural Sciences, National University of Science
%and Technology MISIS (NUST MISIS), \\ Almalyk Branch, Almalyk
%110105, Uzbekistan}

\end{center}

\vskip1.0cm

{\bf Abstract:} 
In this work, we performed an energy-dependent study of low-frequency oscillations observed in GX 13+1 using \textit{AstroSat} (Large Area X-ray Proportional Counter and Soft X-ray Telescope). The hardness-intensity diagram (HID) of the observation resembles a `$\nu$'-shaped track, while the color-color diagram exhibits a `$<$'-shaped track, similar to the horizontal and normal branches of the Z source. We conducted flux-resolved temporal studies focusing on low-frequency variability and divided the HID into five regions: A, B, C, D, and E. Low-frequency quasi-periodic oscillations (QPOs) were detected in Regions A, B, and C. The QPO in Region A has a frequency of $5.06^{+0.54}_{-0.48}$ Hz with a quality factor (Q-factor) of 2.80. In Region B, the QPO was detected at $4.52^{+0.14}_{-0.13}$ Hz with a Q-factor of 5.79, while in Region C, it was observed at $4.70^{+0.62}_{-0.42}$ Hz with a Q-factor of 4.35. The QPO frequencies, Q-factors, and low root-mean-square (rms) values (1.32\%, 1.34\%, and 0.7\%) suggest that these oscillations are Normal Branch Oscillations, similar to those reported in GX 340+0. We modeled the rms and lag of the QPOs using a propagative model, considering variations in blackbody temperature, coronal heating rate, and optical depth. Our findings indicate that the observed QPOs are likely driven by interactions between the corona and variations in the blackbody temperature. 


\vskip1.0cm
	{\section{Introduction}}{\label{INtro}}
	X-ray binaries (XBs) are binary star systems composed of a compact object, either a neutron star (NS) or a black hole (BH), accreting matter from a companion star, typically a main-sequence star, and are prominent X-ray emitters. These systems are classified based on the mass of the donor star: high-mass X-ray binaries (HMXBs), where the donor mass is $\geq 10 M_\odot$, and low-mass X-ray binaries (LMXBs), with donor masses $\leq 1 M_\odot$. A subclass of LMXBs containing a NS, known as low-mass NS X-ray binaries (LM-NSXBs), can be further categorized into atoll and Z sources, distinguished by their spectral and timing properties~\cite{1,2}. In atoll sources, the color-color diagram (CCD) or the hardness-intensity diagram (HID) reveals two primary states: the island state and the banana state. In contrast, Z sources trace a characteristic Z-shaped track in the CCD/HID, consisting of the horizontal branch (HB), the normal branch (NB), and the flaring branch (FB). The spectral characteristics of atoll sources are hard in the island state and soft in the banana state, whereas Z sources generally show soft spectra dominated by blackbody emission. Additionally, atoll and Z sources exhibit distinct variability patterns, with Z sources tending to show variability at lower or softer energies~\cite{3a}. 

	NSXBs display various forms of temporal variability in their power density spectra (PDS), including broadband noise (BBN) and quasi-periodic oscillations (QPOs). In atoll sources, the variability is typically characterized by two main noise components, very low-frequency noise (VLFN) and high-frequency noise (HFN), and occasionally QPOs. The VLFN follows a power-law distribution, described by $P \propto \nu^{-\alpha}$, with a spectral index ranging from 1 to 1.5. The HFN component is modeled by a power law with an exponential cutoff: $P \propto \nu^{-\alpha} e^{-\nu/\nu_{\rm cut}}$, where $0.8 \le \alpha \le 1$ and the cutoff frequency lies between 0.3 and 25 Hz. VLFN is typically observed in the banana state, often accompanied by HFN, while the island state is dominated by HFN and may contain broad features or ``bumps.'' Compared to Z-sources, atoll sources exhibit more pronounced short-term variability~\cite{2}. 

The PDS of Z-sources typically reveal three main BBN components: VLFN, characterized by a power-law index in the range 1.5 $\leq \alpha \leq$ 2; HFN, with $\alpha \sim 0$ and a cutoff frequency between 30 and 100 Hz; and LFN, also with $\alpha \sim 0$ but with a cutoff frequency spanning 2 to 20 Hz~\cite{3}. In addition to these noise components, Z-sources display several types of QPOs, including horizontal branch QPOs (HBOs) with centroid frequencies in the range of 15–60 Hz, normal branch QPOs (NBOs) centered around 5–7 Hz~\cite{11x,11y,11m}, flaring branch QPOs (FBOs) with centroid frequencies between 6 and 20 Hz, as well as kHz QPOs~\cite{3}.

	Atoll sources and Z-sources exhibit several common features, such as the presence of kHz QPOs and the trend of increasing QPO frequency across different spectral branches or states. However, they differ primarily in their mass accretion rates ($\dot{M}$). Specifically, atoll sources typically have $\dot{M} \lesssim 0.5 \dot{M}_{\rm Edd}$, while Z-sources accrete matter at rates close to the Eddington limit ($\dot{M} \sim \dot{M}_{\rm Edd}$). Another key distinction lies in their magnetic field strengths: atoll sources possess magnetic fields weaker than $10^9$ Gauss, whereas Z-sources generally exhibit fields in the range of $10^9$–$10^{10}$ Gauss~\cite{3b}.
	
	\begin{figure}
	\centering
	\includegraphics[width=0.8\linewidth]{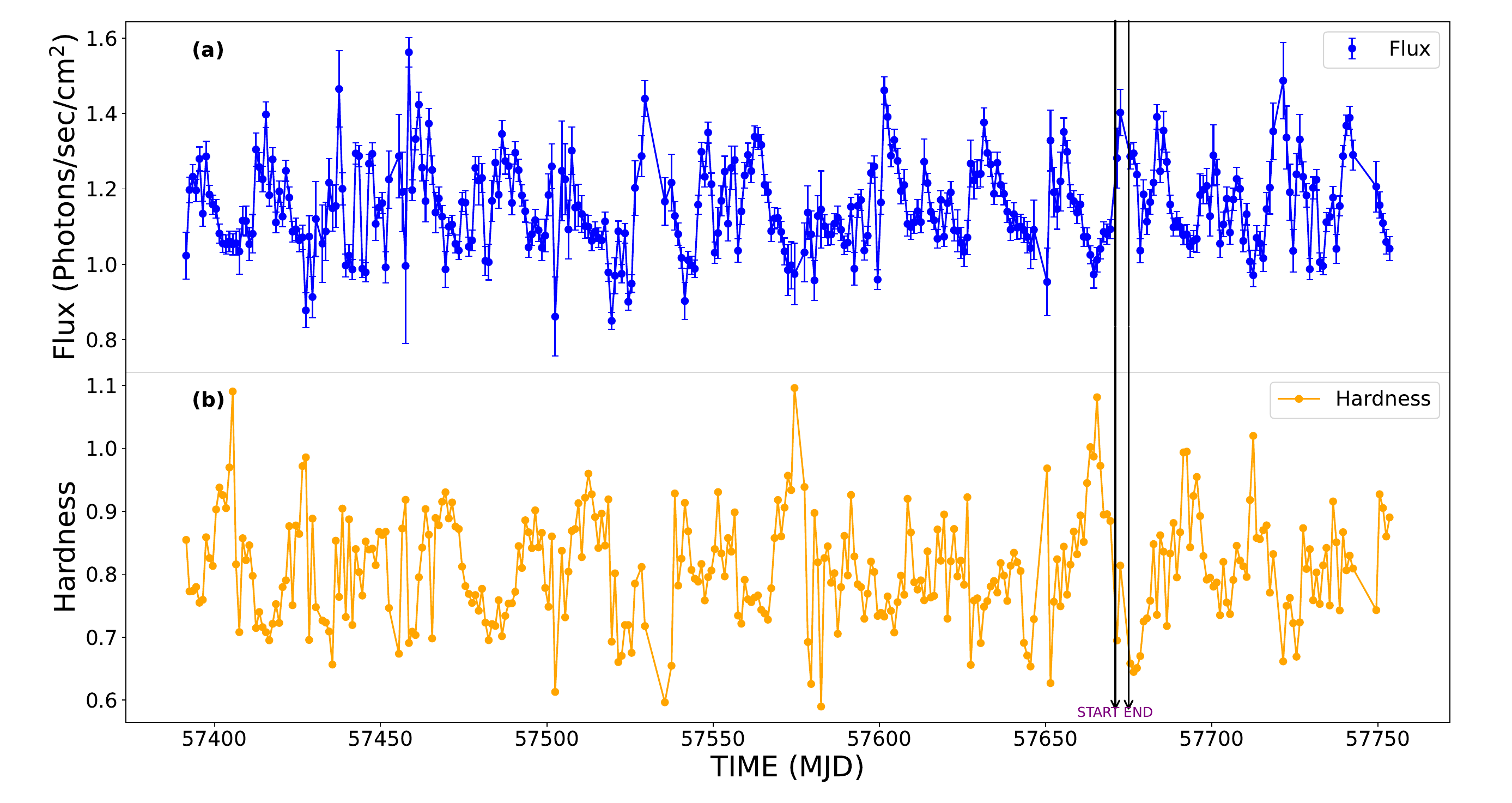}
	\captionof{figure}{MAXI Long-term Light Curve of GX 13+1. (a) Displays the 2-20 keV light curve with a 1-day bin for the year 2016. (b) Shows the evolution of hardness (4-10 keV/2-4 keV) over time (in MJD). The \astrosat{} observations are marked with black arrows, indicating the start and end times.} 
	\label{LC_maxi}
\end{figure}

	To explore the radiative mechanisms responsible for the observed variability in the source, such as QPOs and BBN, it is crucial to examine the energy dependence of fractional rms and phase (or time) lags. One approach to this involves a Comptonization-based model that has been utilized to study the energy-dependent behavior of kHz QPOs. This model incorporates variations in the seed photon flux, driven by high-energy photons that are scattered back onto the soft photon source, thereby creating a feedback mechanism~\cite{32b,32d,32c}. The models developed by Karpouzas et al.~\cite{32e} and Bellavita et al.~\cite{32j} provide explanations for both hard and soft lags in low-frequency QPOs, attributing them to thermal reprocessing in the accretion disk and corona. Lyubarskii~\cite{6} suggested that such lags could originate from perturbations propagating inward within the accretion flow. Furthermore, studies by Jithesh et al.~\cite{32a}, Maqbool et al.~\cite{16h}, and Mudambi et al.~\cite{32f} proposed that energy-dependent variability can result from fluctuations propagating from the disk to the corona. These delayed perturbations influence the disk emission and subsequently modulate the coronal heating rate. Following a similar approach, Garg et al.~\cite{12-a} introduced a formalism to interpret the energy-dependent fractional rms and time lags associated with such variability. More recently, this framework was employed by Bhattacherjee et al.~\cite{9,9x} to analyze the energy-dependent timing properties of a neutron star X-ray binary (NSXB). Additionally, Sudha et al.~\cite{11k} investigated frequency- and energy-dependent lags in Cyg X-2 to gain insight into the nature of the detected NBO.

	Galactic X-ray source (GX) 13+1 is a persistent bright LM-NSXB at a distance of 7$\pm$1 kpc in the Galactic bulge with the companion star K5 III star with mass M$_{\rm donor}$ = $\rm 5 M_\odot$~\cite{3c,3d}. The binary orbital period according to the infrared and X-ray light curve is reported to be 24.5 days~\cite{3e}, and it has an inclination of 60$^\circ$-80$^\circ$~\cite{3f}. 
\begin{figure}
	\centering
	\includegraphics[width=0.5\linewidth, angle=-90]{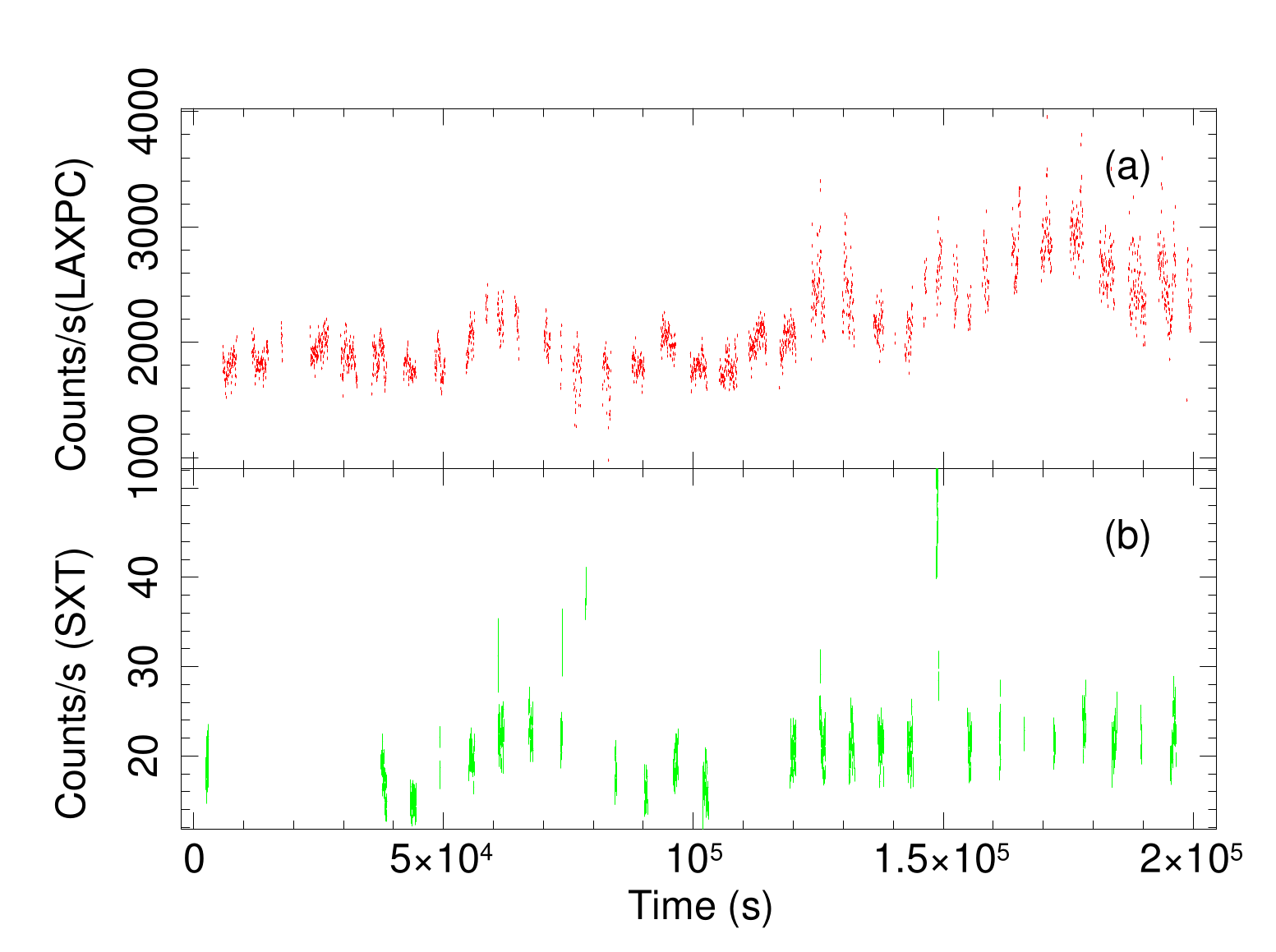}
	\captionof{figure}{ Light curve of the source with a time bin of 32 s, observed using LAXPC (3.0–80.0 keV; panel a) and SXT (0.3–8.0 keV; panel b).}
	\label{LC}
\end{figure}
	
	GX 13+1 is an interesting source because it was initially considered to be one of the bright GX atoll sources~\cite{1,3f}, but it is now classified as a Z-source due to its evolution in the CCD/HID~\cite{3i}. Like other bright GX sources, such as GX 3+1, type I bursts~\cite{11,16k} have been reported in this source, first in 1985 with the \emph{SAS-3} mission~\cite{3g} and then in 1995 with \emph{Ginga}~\cite{3h}, which confirmed the presence of NS as a compact object. QPOs similar to HBO have been reported in GX 13+1 in a range of 57-69 Hz, along with HFN at a cut-off frequency of $\sim$2 Hz~\cite{3b}. Similarly, Schnerr et al.~\cite{3j} reported 57-69 Hz QPOs using \rxte{} observations. Giridharan et al.~\cite{3k} also reported QPOs in the 50-60 Hz range using \astrosat{}. Periodic X-ray absorption dips have been reported, suggesting that it is a dipping source, which occurs when X-rays from the central parts are occulted by the outer parts of the accretion disk~\cite{3l}. As a dipping source, various absorption edges and lines have been reported with various mission observations, such as \emph{ASCA}~\cite{3m}, \emph{Chandra}~\cite{3p}, and \xmm{}~\cite{3r} observations. Saavedra et al.~\cite{3s} discovered evidence of relativistic reflection in the Fe K line, as well as in the Compton hump at 10-25 keV using \nustar{}. Giridharan et al.~\cite{3k} also reported the presence of a positive residual at around 6.5 keV, referring to the Fe K X-ray fluorescence line.

 In Garg et al.~\cite{12-a}, the authors discuss a propagative model that considers fractional rms and phase lag/time lag of the peak frequency (either QPO or BBN) observed in the PDS. This model describes the energy-dependent nature of the variability and explains its origin. The authors verified the effectiveness of their model using the BH X-ray binary (BHXB) GRS 1915+105, fitting the model to the millisecond lag observed during the observation and explaining the variations in spectral parameters due to perturbations in the system. They found that the perturbation originated in the inner disk and propagated toward the coronal region, causing parameter variations due to the viscous or sound-crossing time scale. They also applied this technique to MAXI 1535-571\cite{16g}, where they considered $|\delta kT_{in}|$ as the variation in the accretion rate $|\delta \dot{M}|$ and the variation in the disk normalization $|\delta N_{dbb}|$ as the ratio of the inner radius variation to accretion rate variation, $|\delta R_{in}|/ |\delta \dot{M}|$. Similarly, for sources such as H 1743-322~\cite{8}, 4U 1608-52~\cite{9}, and GX 339-4~\cite{17k}, the origin of variability observed in these systems has been satisfactorily explained using the Garg et al.~\cite{12-a} formalism.

GX 13+1 is classified as a Z-track source, and we are interested in investigating the low-frequency variability of the source using the single \astrosat{} observation of GX 13+1.	Since the Large Area X-ray Proportional Counter (LAXPC) has one of the best timing resolutions for performing temporal studies, we decided to revisit this observation to understand the characteristics of low-frequency quasi-oscillations or BBN by employing the propagative model proposed by Garg et al.~\cite{12-a}. 

The manuscript is structured as follows: in Section 2, the data reduction procedure and detailed broadband spectral and temporal analyses are explained. In Section 3, we discuss the results of the spectral and temporal analyses, along with the model fit to the rms and time lag. Finally, in Section 4, we present the conclusions of our findings.

	{\section{Observation and data analysis}{\label{ODA}}}
	\subsection{Observation and data reduction}{\label{ODR}}

	\astrosat{}~\cite{12-b} observed GX 13+1 from October 10, 2016, to October 14, 2016 (Observation ID G06 032T02 9000000730). Figure \ref{LC_maxi} (a) illustrates the MAXI long-term light curve for GX 13+1 for the entire year 2016, as well as the hardness (4-10 keV/2-4 keV) evolution over time in Modified Julian Date (MJD) in (b). The two arrows in Figure \ref{LC_maxi} indicate the start and end times of the \astrosat{} observation. The LAXPC~\cite{12-c} and Soft X-ray Telescope (SXT)~\cite{13,14} of \astrosat{} simultaneously observed GX 13+1 with exposures of 79.06 ks and 24.97 ks, respectively. The LAXPC data were reduced using the \texttt{LAXPCsoftware}\footnote{\url{http://astrosat-ssc.iucaa.in/laxpcData}}(\texttt{LAXPCSOFT}; version 2022 August 15). Using standard sub-routines of LAXPC analysis available in \texttt{LAXPCsoftware}\footnote{\url{http://astrosat-ssc.iucaa.in/laxpcData}}, we extracted the light curve and spectra. The Level 1 data of the SXT observation were obtained from the \texttt{ISSDC} website\footnote{\url{https://webapps.issdc.gov.in/astro_archive/archive/Search.jsp}}, and the Level 2 data were obtained by running the SXT pipeline – AS1SXTLevel2-1.4b\footnote{\url{https://www.tifr.res.in/~astrosat_sxt/sxtpipeline.html}} on the Level 1 data. 
\begin{figure}
	\begin{center}
		$%
		\begin{array}{c@{\hspace{.1in}}cc}
			\includegraphics[width=3.3 in, height=2.5 in]{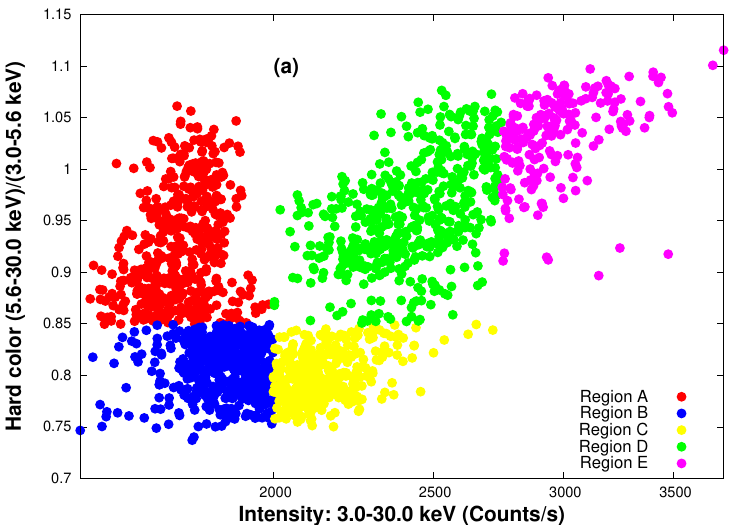} &%
			\includegraphics[width=3.3 in, height=2.5 in]{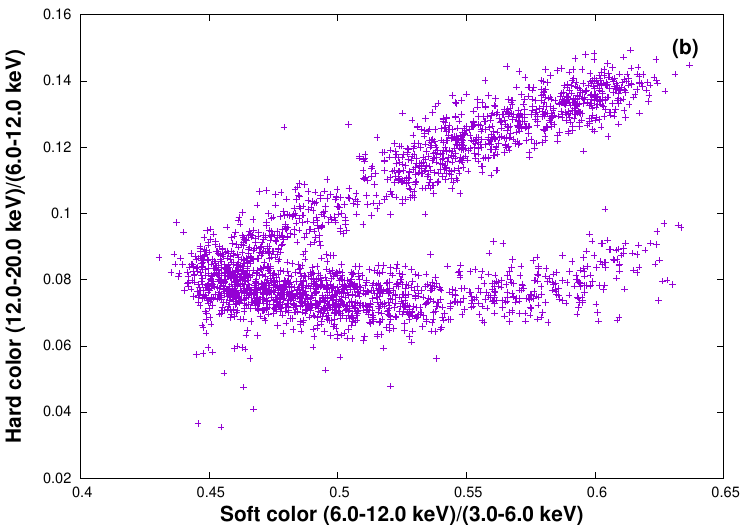} \\  

		\end{array}%
		$%
	\end{center}
	\caption{\label{HID_CCD} The HID (a) and CCD (b) of the source using LAXPC data with a time bin of 32 seconds. The HID illustrates different flux states used for performing flux-resolved spectroscopy, while the CCD displays the characteristic ``$<$" pattern commonly observed in GX 13+1.}

\end{figure}
	The source count rate in SXT was $\sim$ 18.22 counts/sec in the fast windowed photon counting mode (FW). Hence, there was no issue with pile-up, as the threshold value in FW mode is very high (i.e., photon counts/sec $>$ $\sim$340 counts s$^{-1}$). The merging of individual orbits was carried out using the SXT event merger tool\footnote{\url{https://www.tifr.res.in/astrosat_sxt/dataanalysis.html}}, and a clean event file was also extracted. Using the \textbf{HEASARC} (version 6.32.1) tool \texttt{XSELECT}, we checked the counts and extracted the light curve and spectra. We corrected the auxiliary response file (ARF) for the vignetting effect using the \texttt{sxtARFModule} tool\footnote{\url{https://www.tifr.res.in/astrosat_sxt/dataanalysis.html}} and grouped the photon spectra with the corrected ARF, response matrix file (RMF) \texttt{``sxt pc mat g0to12.rmf''}, and blank sky background file \texttt{(SkyBkg comb EL3p5 Cl Rd16p0 v01.pha)} provided by the SXT payload operation center (POC). The extracted light curves from LAXPC and SXT, with a time bin of 32 seconds, are shown in Figure \ref{LC} (a) and (b), respectively.

	\subsection{Data analysis}
	
	\subsubsection{Timing Analysis}{\label{TA}}

	\begin{figure}[h!]
		\centering
		\includegraphics[width=0.45\linewidth]{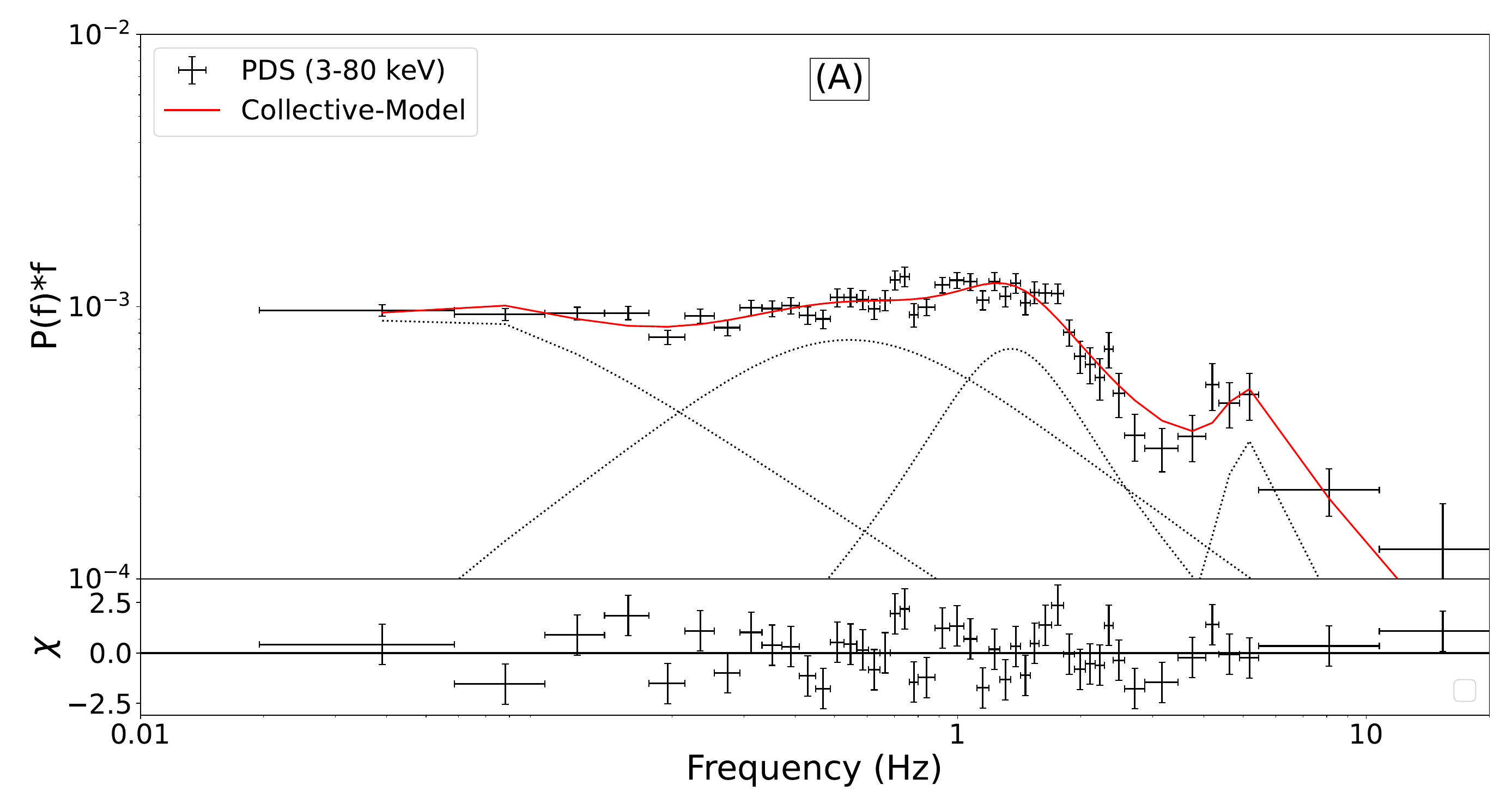}
		\includegraphics[width=0.44\linewidth]{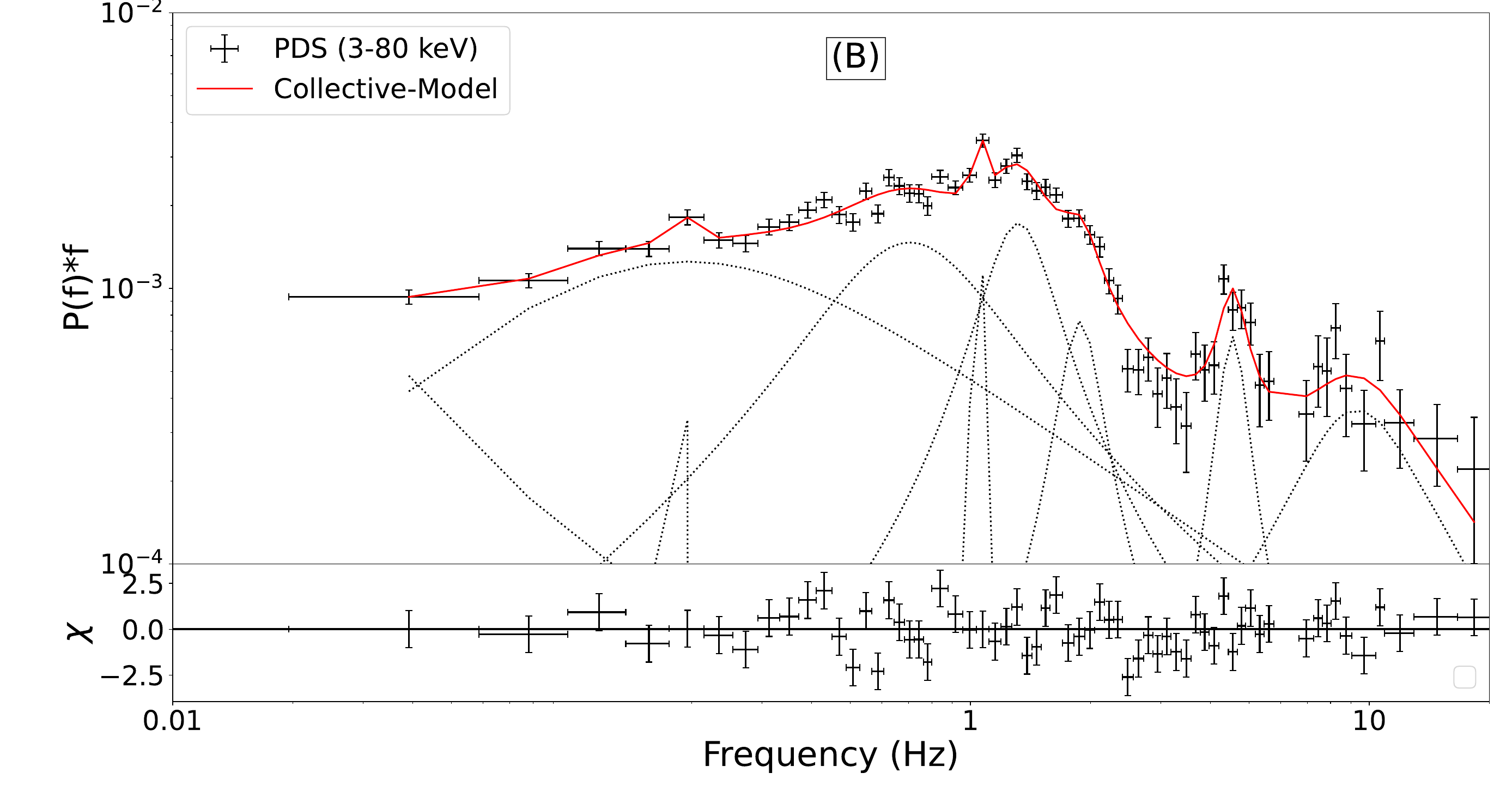}
		\includegraphics[width=0.45\linewidth]{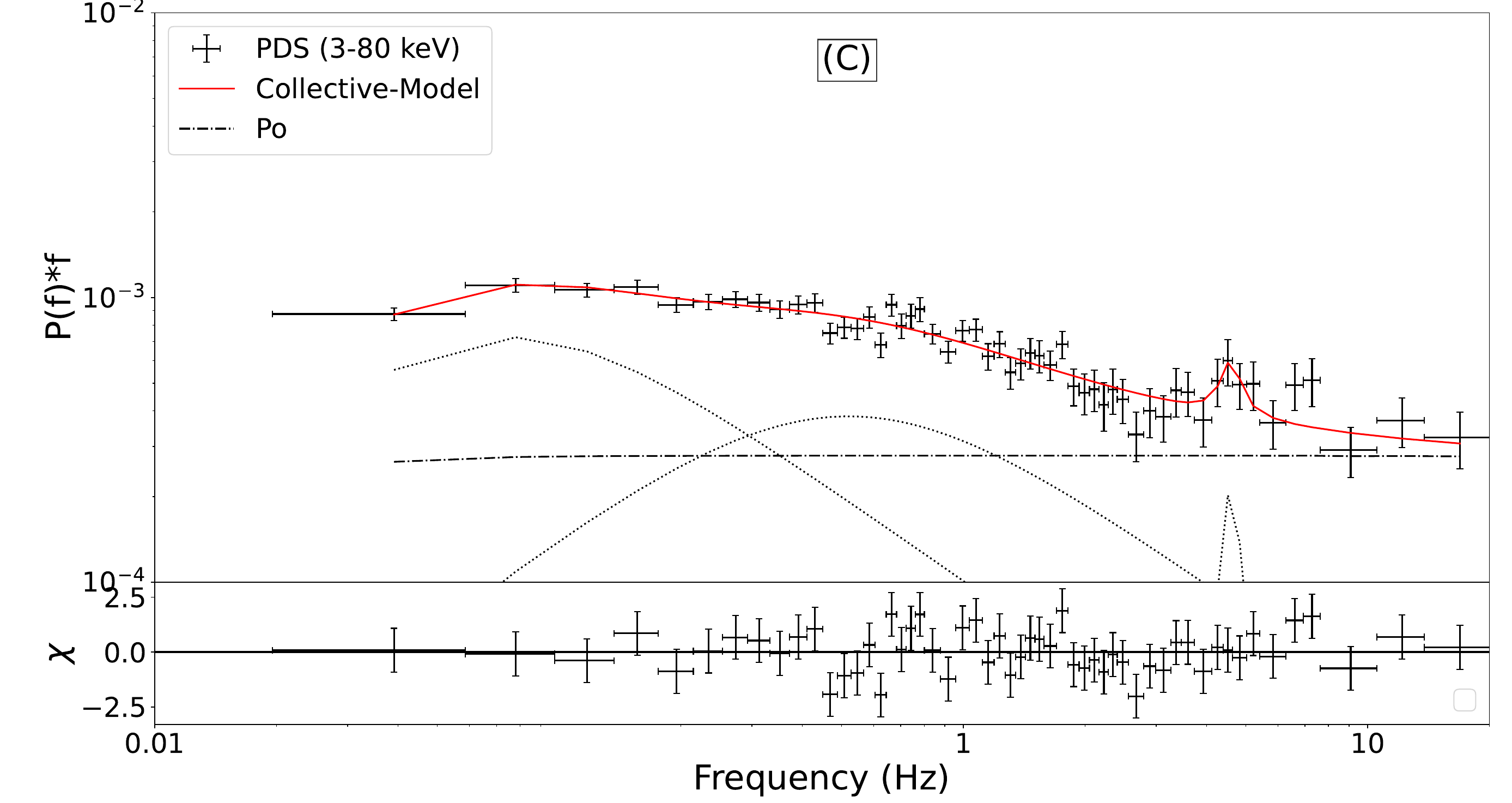}
				\includegraphics[width=0.44\linewidth]{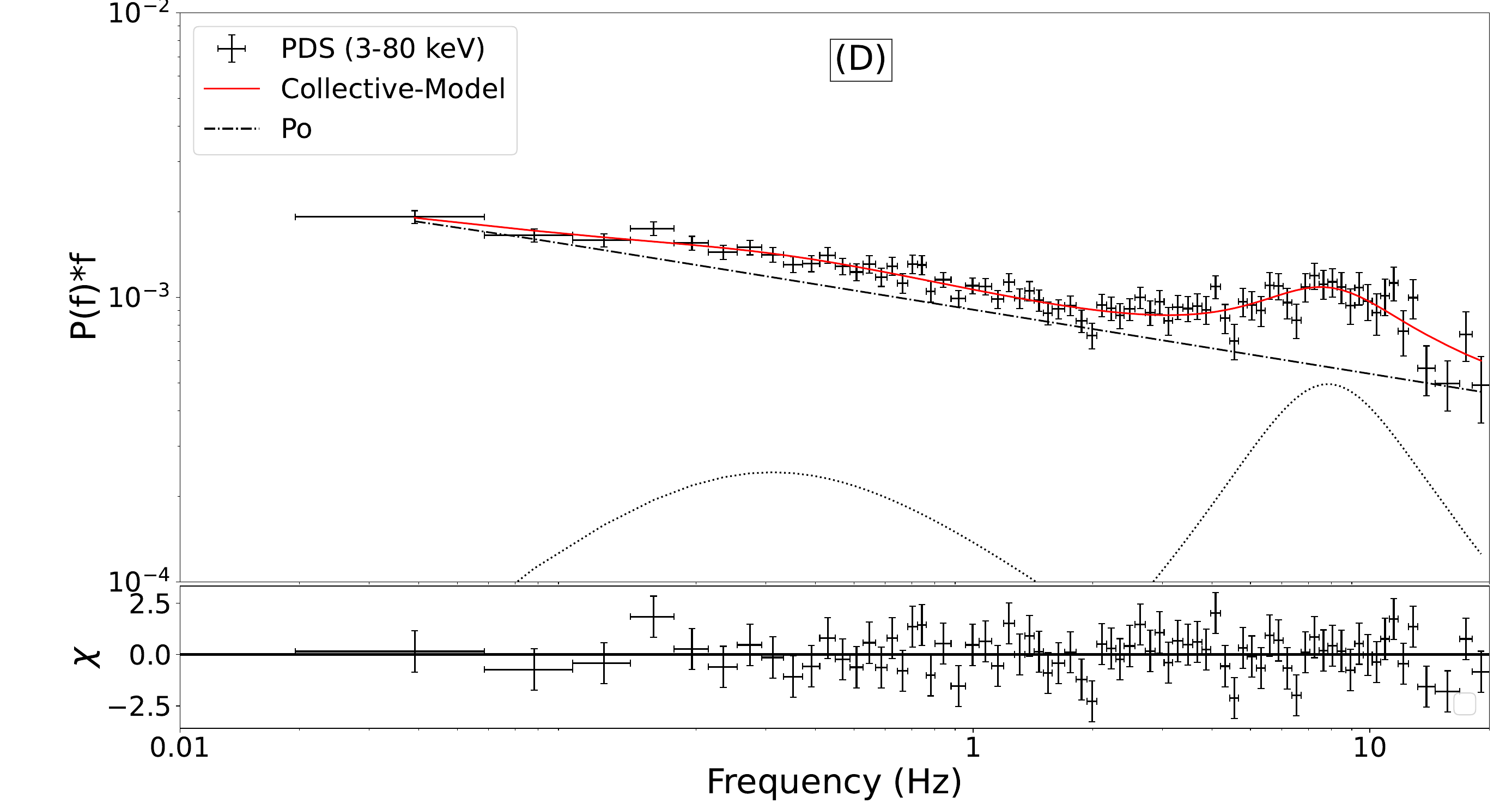}
						\includegraphics[width=0.5\linewidth]{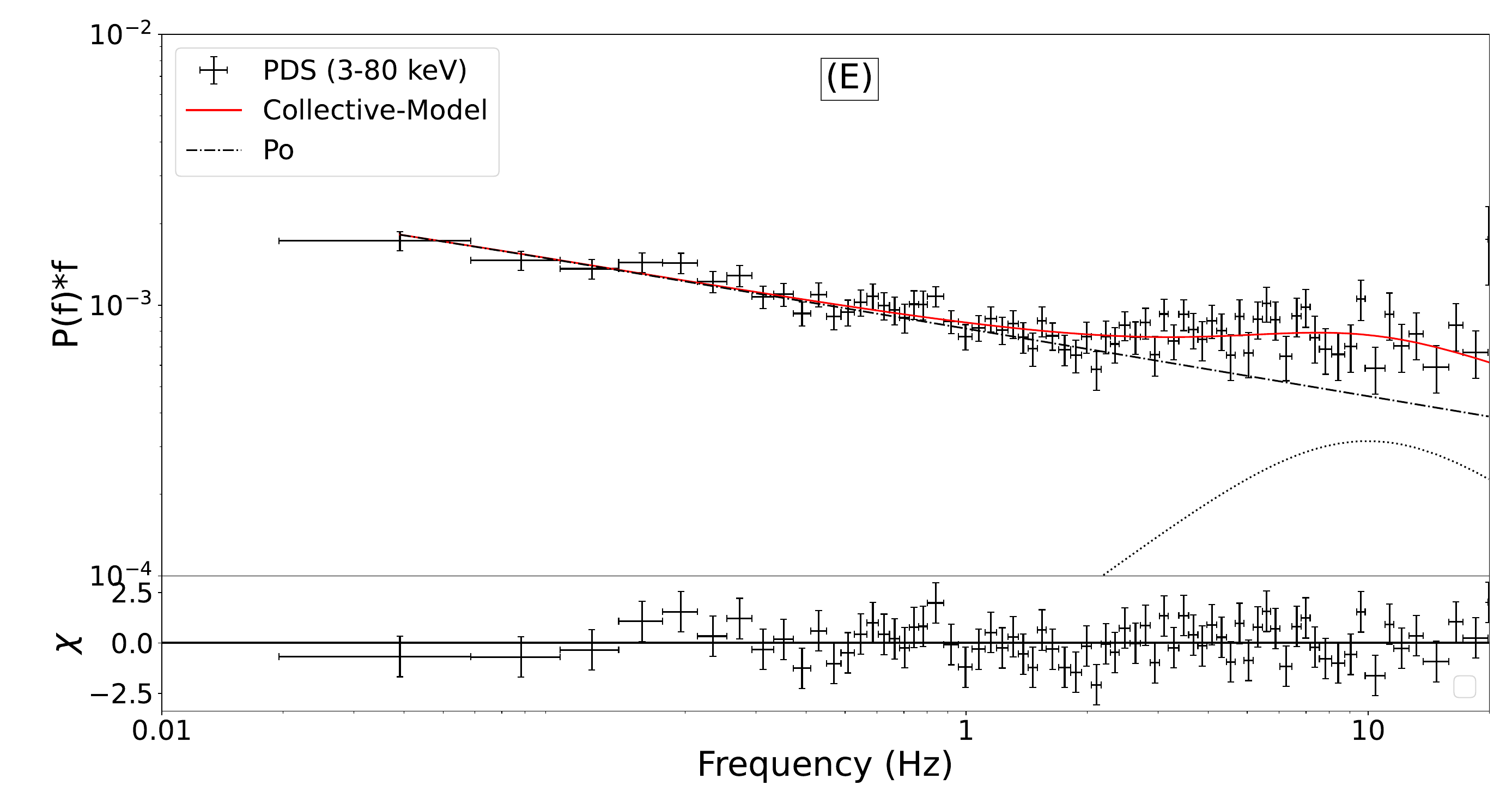}
		\captionof{figure}{
			The PDS of different regions, fitted in an energy band of 3-80 keV, is fitted with a power law (black dot-dashed line) and multiple Lorentzians (black dotted lines) in a frequency band of 0.05-20 Hz.}
		\label{PDS_fit}
	\end{figure}
	
It can be clearly seen that the CCD and HID (Figure \ref{HID_CCD}) show NB and HB similar to what is observed in Allen et al.~\cite{12}. Although HB appears similar to FB, we consider it as HB due to the presence of a HBO, a type of QPO, as reported by Giridharan et al.~\cite{3k}. PDS of the energy band of 3-80 keV within the frequency range of 0.05 to the Nyquist frequency of 20 Hz were generated using the \texttt{laxpc\_find\_freqlag} subroutine for all regions. The generated PDS has Poisson noise subtracted, using the default dead time value of 42 $\mu$s ~\cite{11z}.
	
We modeled the extracted PDS using multiple Lorentzians, including one or two zero-centered Lorentzians, as shown in Figure \ref{PDS_fit}. The collective model, shown in red, represents all the contributions of the Lorentzians. In some regions, we also used a power law to account for faint noise. Multiple Lorentzian functions are commonly employed to model various components in the PDS, with each Lorentzian described by the following expression \ref{lor_eqn}.:- \\
		\begin{equation}
			L(\nu) = \frac{r^{2} \Delta}{\pi} \frac{1}{(\nu - \nu_{0})^2 + \Delta^{2}}
			\label{lor_eqn}
		\end{equation}\label{EQ1}\\
 In this formulation, `$r$' denotes the integrated rms amplitude of the component across the full frequency range, $\nu_{0}$ is the centroid frequency and `$\Delta$' is the half-width at half-maximum (HWHM). The characteristic or peak frequency of the Lorentzian, representing the frequency at which the component's contribution is most significant, is given by $\nu_{max}$ = $\sqrt{{\nu_{0}}^{2}+ \Delta^{2}}$ ~\cite{11w}.
		\\
	We segmented the HID (Figure \ref{HID_CCD}a) into five regions as follows: (i) Region A: 0.85–1.10 (1500–2000 counts/s), (ii) Region B: 0.72–0.85 (1500–2000 counts/s), (iii) Region C: 0.75–0.85 (2000–2750 counts/s), (iv) Region D: 0.85–1.10 (2000–2750 counts/s), and (v) Region E: 0.85–1.15 (2750–3800 counts/s). These regions are almost exactly the same as the flux regions defined by Giridharan et al. ~\cite{3k}. We detected the presence of low-frequency oscillations in Regions A, B, and C, which can be considered QPOs based on their Q-factor values (the ratio of the characteristic frequency to the full width at half maximum (FWHM)), Q=$\nu$/FWHM and the significance as clearly shown in Figure \ref{PDS_fit} ((A), (B) and (C)). In Region A, we detected a QPO at $\sim$ 5.06 Hz with a Q-factor of $\sim$ 2.80. In Region B, we detected a QPO at $\sim$ 4.52 Hz with a Q-factor of $\sim$ 5.80, while in Region C, we detected a QPO at $\sim$ 4.58 Hz with a Q-factor of 8.30. The significance of these QPOs was $\lesssim$3. This frequency range and the significance level of these features indicate that QPOs are similar to NBOs usually seen in Z-track sources (such as GX 340+0 \cite{11x,11y}). 
		
		The QPOs detected at $\sim$ 5 Hz in this observation could be the first report of such QPOs in this system. The QPOs in the frequency range of 50-60 Hz have been reported previously in this source ~\cite{3b,3k} which were referred to as the HBO. 
		
		The PDSs of regions D and E were modeled using a Lorentzian and a power-law component. Low-frequency oscillations, such as those observed in Regions A, B, and C, were not present. The PDS fitted with a power law (black dot-dashed line) and a Lorentzian is shown in Figures \ref{PDS_fit}(d) and \ref{PDS_fit}(e). 
	
	\subsubsection{Spectral Analysis}{\label{SA}}
	We have tried to reproduce the spectral result of the Giridharan et al.~\cite{3k} by following the exact procedure as mentioned in their work. We performed the flux resolved analysis as similar to that of Giridharan et al.~\cite{3k}. We were able to reproduce the similar values of the parameters as reported.
		Our primary goal is to employ the method of Garg et al. ~\cite{12-a}, where spectral information will be used to determine physical parameters to fit the energy-dependent rms and time lag. 
		We will use the spectral parameters to model the fractional rms and lag. The process of extracting fractional rms and lag is discussed in the upcoming subsection \ref{rms_lag}. The spectral parameters of \texttt{thcomp} ~\cite{4,16a} and \texttt{bbodyrad} will aid in determining these physical parameters. Further discussion will be carried out in subsection \ref{Energy-dep}.

\subsubsection{Fractional rms and lag generation.}{\label{rms_lag}}
The rms and time lag in the 3–30 keV energy band were computed using the LAXPC subroutine \textit{laxpc$\_$find$\_$freqlag}, with a signal-to-noise ratio ($s$) threshold of 3. This subroutine requires two primary input parameters: frequency resolution ($\Delta$$f$) and central or characteristic frequency ($f$) of the targeted feature. To determine the fractional rms, the routine integrates the power spectrum over the frequency interval from $f-\Delta f$ and $f+\Delta f$ for the selected energy band and then takes the square root of the result. The frequency resolution $\Delta f$ is typically chosen to be either equal to or half the full width at half maximum (FWHM) of the broad frequency component for which the lag is being estimated.

The characteristic frequency corresponding to the QPO frequency for extracting the rms and lag was calculated using equation \ref{CHA_1212},
\begin{equation}
	f = \sqrt{\nu_0^2 + \left(\frac{\Delta}{2}\right)^2}
	\label{CHA_1212}
\end{equation}
where:
\begin{itemize}
	\item $f$ is the characteristic frequency,
	\item $\nu_0$ is the centroid frequency,
	\item $\Delta$ is the FWHM.
\end{itemize}
Equation \ref{CHA_1212} is widely used in dealing with BBN whereas in the QPO frequency scenario the characteristic frequencies comes out in the similar values to the QPO frequency. Here, we have used 5.06 Hz in Region A, 4.52 Hz in Region B, and 4.58 Hz in Region C as `$f$' to extract rms and lag.

The subroutine also calculates the phase lag of the cross-spectrum between two light curves over distinct energy bands, employing one as a reference band. The phase lag thus generated can be treated as a time lag by dividing $2\pi f$~\cite{11p}.
 
We considered the reference energy band to be 3-6 keV. We generated the rms and lag for the QPO frequency/centroid frequency in the energy range of 3-30 keV. In Figure \ref{RMS_LAG} (a), (c), and (e), we observed a gradual increase in rms along with an increase in energy for all regions. We observed a very marginal variation in the lag profile in Regions A and C across some energy bands, whereas the lag in Region B exhibited significant variations. In contrast, Regions D and E showed no variation in the lags.
 
The time lag of Region A is quite consistent with 0, while the lag of Region B is significant for several energy bins. In Region C, the lag is consistent with 0 after the third energy band, as shown in Figure \ref{RMS_LAG} (f). For Region B, we observed an initial positive lag of up to 30 ms for the energy band 12.95-18.5 keV, and in the next energy band, the lag became zero, as shown in Figure \ref{RMS_LAG} (d). In Region C, we observed that the time lag was positive, up to 40 ms, for the energy band 9.25-11.95 keV, but after that it showed a negative lag with marginal variation that is also consistent with 0, as shown in Figure \ref{RMS_LAG} (f). Taking into account the statistical uncertainties in the coherence and lag values, we modeled the lags for all three regions. This lag modeling is meaningful in the context of our analysis, as we are modeling both the energy-dependent fractional rms and the time lag. 

\begin{figure}[!h]
		\centering
		\includegraphics[width=0.3\linewidth]{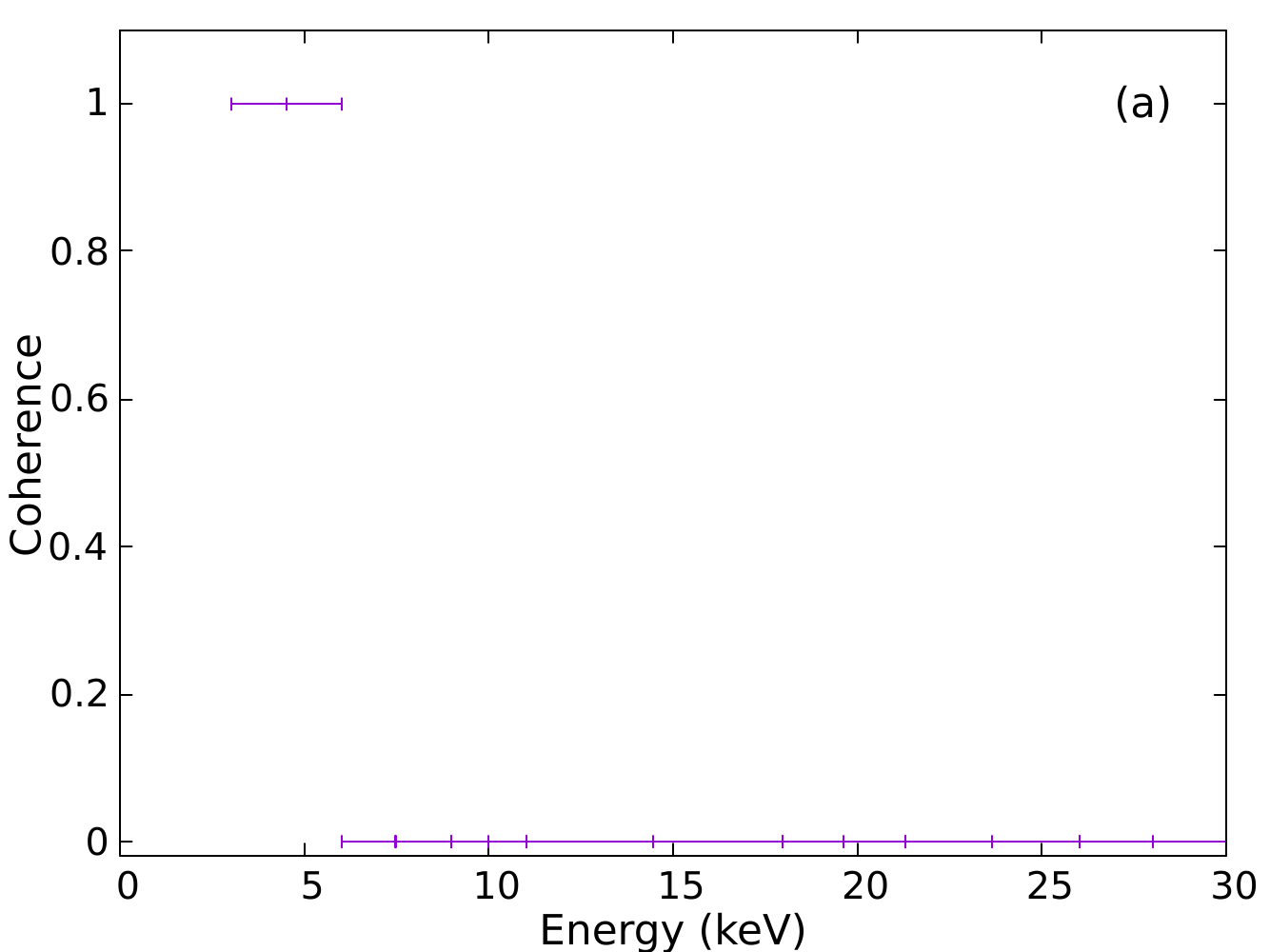}
		\includegraphics[width=0.3\linewidth]{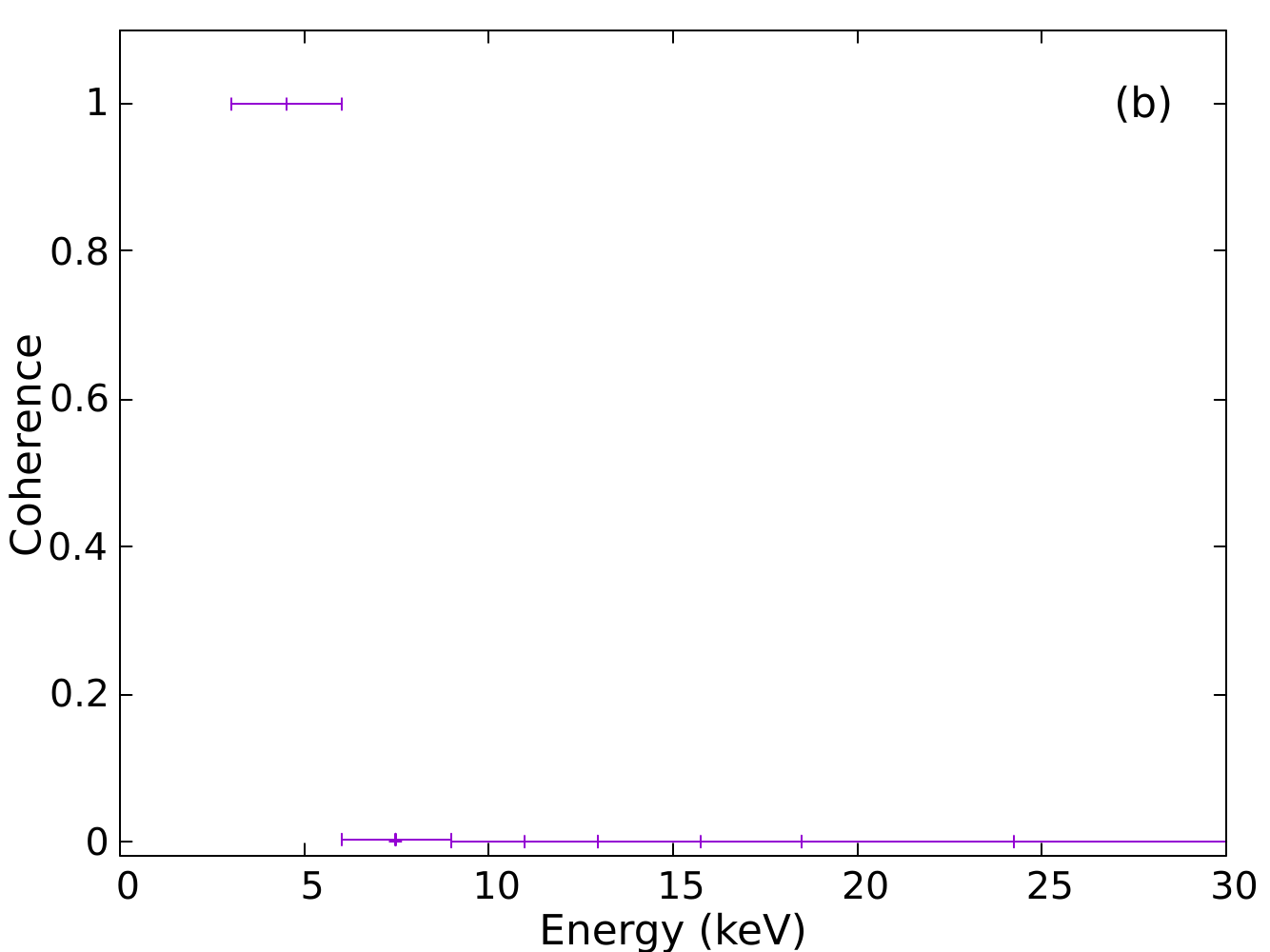}
		\includegraphics[width=0.3\linewidth]{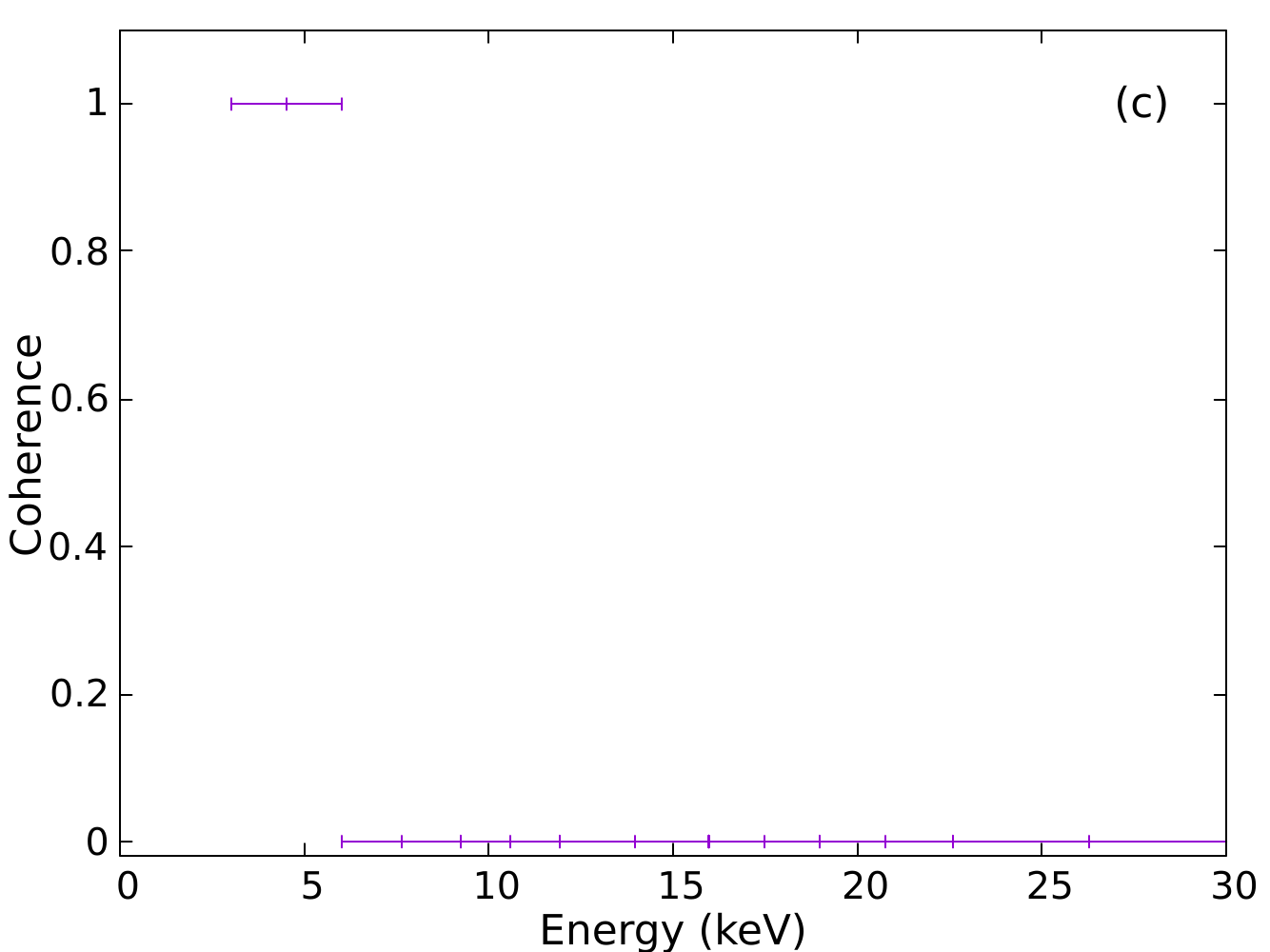}
		\caption{ The coherence as a function of energy for the lags of Region A (a), Region B (b), and Region C (c).}
		\label{Coherence}
	\end{figure}

To check the linear correlation between variability in two energy bands, we computed coherence as a function of energy for the lag spectra of Regions A, B, and C, as shown in Figure \ref{Coherence} (a), (b), and (c), respectively. A sharp decrease in coherence is evident after the reference band. Nowak et al. \cite{11p} observed that at large energy separations, coherence drops below 0.1, and the corresponding lags become unreliable due to noise dominance. When the coherence value is less than 0.1, the corresponding lags are statistically unreliable \cite{11p,12X}. In Figure \ref{Coherence} (a), (b), and (c), although coherence does not show a significant trend, the lag profiles of Regions B (Figure \ref{RMS_LAG}d) and C (Figure \ref{RMS_LAG}f) exhibit notable variations up to certain energy bands. Our goal in modeling the energy-dependent rms and lag, using the formalism by Garg et al. \cite{12-a}, is to understand the origin of the variability. Although the obtained lags exhibit low coherence, rendering them unreliable, we proceeded with modeling to explore whether the model could qualitatively reproduce the observed lag trends and rms variability. When interpreting the results of lag modeling in such cases, caution is necessary, as low-coherence lags are typically indicative of noise-dominated signals, thus compromising their reliability. This interpretation is further supported by our coherence analysis, which revealed no significant trends across any region, including Regions D and E.

 We checked the lag for Regions D and E, but did not observe any variation, as shown in Figure \ref{RMS_LAG} (h). Hence, we performed a further analysis using the rms and lag of Regions A, B, and C.

	\subsubsection{Energy-dependent temporal analysis}{\label{Energy-dep}}
	
In order to investigate the radiative processes responsible for the production of variability in the PDS, we examine the energy-dependent rms and time lag and fit with the propagative model proposed by Garg et al.~\cite{12-a,16g}. The energy dependence of variability can be investigated using methods in which one or more spectral parameters are modulated simultaneously, with phase lags introduced between them~\cite{16h,12-a}. Spectral models in XSPEC such as \texttt{thcomp}, \texttt{diskbb}, and \texttt{bbodyrad} include parameters that can be associated with physical quantities. For instance, in the \texttt{thcomp} model, the asymptotic power-law index $\Gamma$ is related to the optical depth $\tau$ via the following relation~\cite{4,16j}:

\begin{equation} \Gamma = \left[\frac{9}{4} + \frac{3m_ec^2}{kT_{\rm e}((\tau + 3/2)^2 - 9/4)}\right]^{1/2} - \frac{1}{2} \label{eq:gam} \end{equation}

Here, $kT_{\rm e}$ represents the electron temperature. The coronal heating rate $\dot{H}$ can also be estimated using $kT_{\rm e}$ (see~\cite{12-a} for further details), where $\dot{H}$ is defined as the energy difference between the Comptonized photons and the original seed photons. As demonstrated by Garg et al.~\cite{12-a,16g}, temporal variations in spectral parameters can induce changes in the steady-state spectrum $F(E)$, which can be expressed as:

\begin{equation} \Delta F (E) = \sum_{j=1}^M \frac{\partial F (E)}{\partial \alpha_j} \Delta \alpha_j \label{DeltaSE} \end{equation}

In this equation, $\alpha_j$ denotes the $j^{\text{th}}$ physical parameter among a total of $M$ parameters, and $\Delta \alpha_j$ represents its variation, typically considered a complex quantity. The rms variability as a function of energy is then given by $(1/\sqrt{2})|\Delta F(E)|/F(E)$. Moreover, the phase lag between photons at energy $E$ and a reference energy $E_{\rm ref}$ is derived from the argument of the complex product of $\Delta F(E)$ and the complex conjugate of $\Delta F(E_{\rm ref})$, as detailed in Garg et al.~\cite{12-a,16g}.	
	
	This stochastic propagation model proposed by Garg et al.~\cite{12-a,16g} has been widely applied to BHXBs (~\cite{9XX,8,17k}). Furthermore, the scheme of Garg et al.~\cite{12-a} has also been tested on LM-NSXBs. Bhattacherjee et al.~\cite{9,9x} applied the scheme to understand the behavior of observed rms and lags of variability in 4U 1608-52 and GX 9+9, respectively. Note that to account for the propagation of variations between the NS surface/boundary layer and the corona in LM-NSXBs, the authors of Garg et al.~\cite{12-a} updated their model's code by replacing the \texttt{diskbb} model with a blackbody model like \texttt{bbodyrad}.

	\begin{table*}[htb!]
		\centering
		\caption{The best-fit parameters of ``MODEL 2'' fitted to the rms and time lag of Region A, Region B, and Region C, calculated at the QPO frequency ($\nu_{\rm QPO}$).}
		\label{S1_S2}
		\scalebox{1.0}{  % Adjust the table width to fit the page width
			\begin{tabular}{@{}ccc ccc@{}}
				\hline 
				Regions & $\nu_{\rm QPO}$& $\delta kT_{\rm bb}$ & $ \delta_{\tau}$  & $\delta \Dot{H}$ &  $\chi^2$/dof \\ \hline\hline \\
				A & 5.06 &$0.004^{+0.001}_{-0.001}$ & $0.12^{+0.08}_{-0.08}$ & $0.02^{+0.010}_{-0.014}$  &  4.96/9 \\ \\
				B & 4.52 &$0.005^{+0.001}_{-0.001}$ & $0.07^{+0.03}_{-0.03}$ & $0.012^{+0.004}_{-0.004}$  & 9.12/8\\ \\
				C & 4.58 &$0.003^{+0.001}_{-0.001}$ & $0.06^{+0.04}_{-0.03}$ & $0.01^{+0.005}_{-0.005}$ & 9.78/8\\ \hline
			\end{tabular}
			
		}
		\\	Note: dof denotes the degree of freedom
	\end{table*}
	
	The generated rms and lag were fitted with a model consisting of the physical parameters. The spectral parameters obtained from the spectral fitting were used to convert into physical parameters. We primarily explored the possibility of variation in the blackbody temperature and the coronal parameters. We started by fitting the Region A rms and lag spectra, considering a very simple scenario with the variation of boundary layer temperature ($|\delta kT_{\rm bb}|$) and heating rate variation ($|\delta \dot{H}|$), with $\dot{H}$ varying with phase lag ($\phi_{\rm \dot{H}}$) w.r.t. $kT_{\rm bb}$. We denoted this as MODEL 1. However, MODEL 1 did not provide a good fit, as the reduced $\chi^2$ was greater than 2 ($\chi^2_{\nu} > 2$). We further introduced variations in parameters, keeping a maximum of five free parameters (three from MODEL 1 and two newly introduced variations), such as the variation in $\tau$ ($|\delta \tau|$) and $\tau$ varying with a phase lag ($\phi_{\tau}$) w.r.t. $kT_{\rm bb}$. We denoted this as MODEL 2, which gave a better and acceptable fit of $\chi^2_{\nu} \sim 0.55$. 
		
\begin{figure}[!h]
		\centering
		\includegraphics[width=0.45\linewidth]{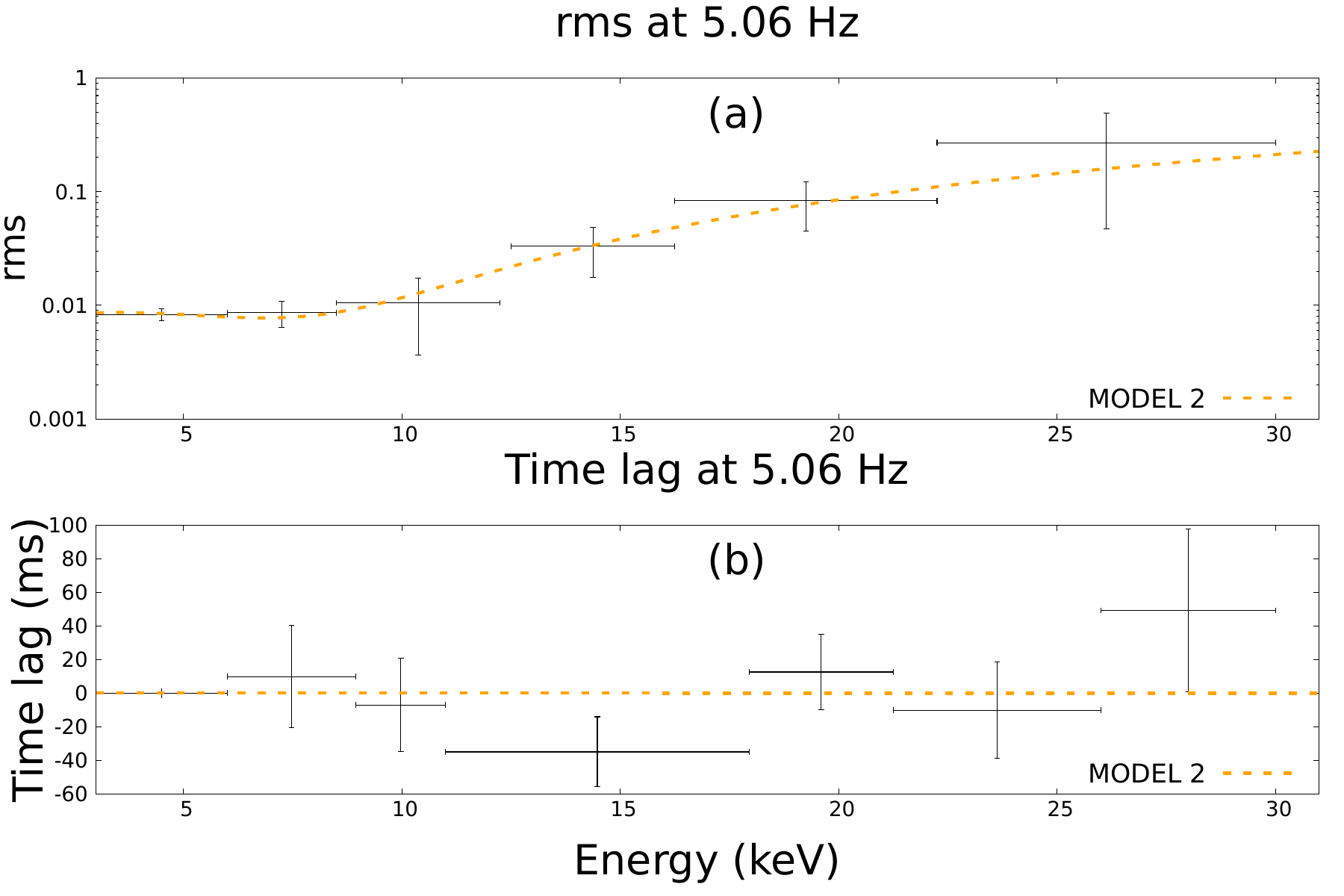}
		\includegraphics[width=0.45\linewidth]{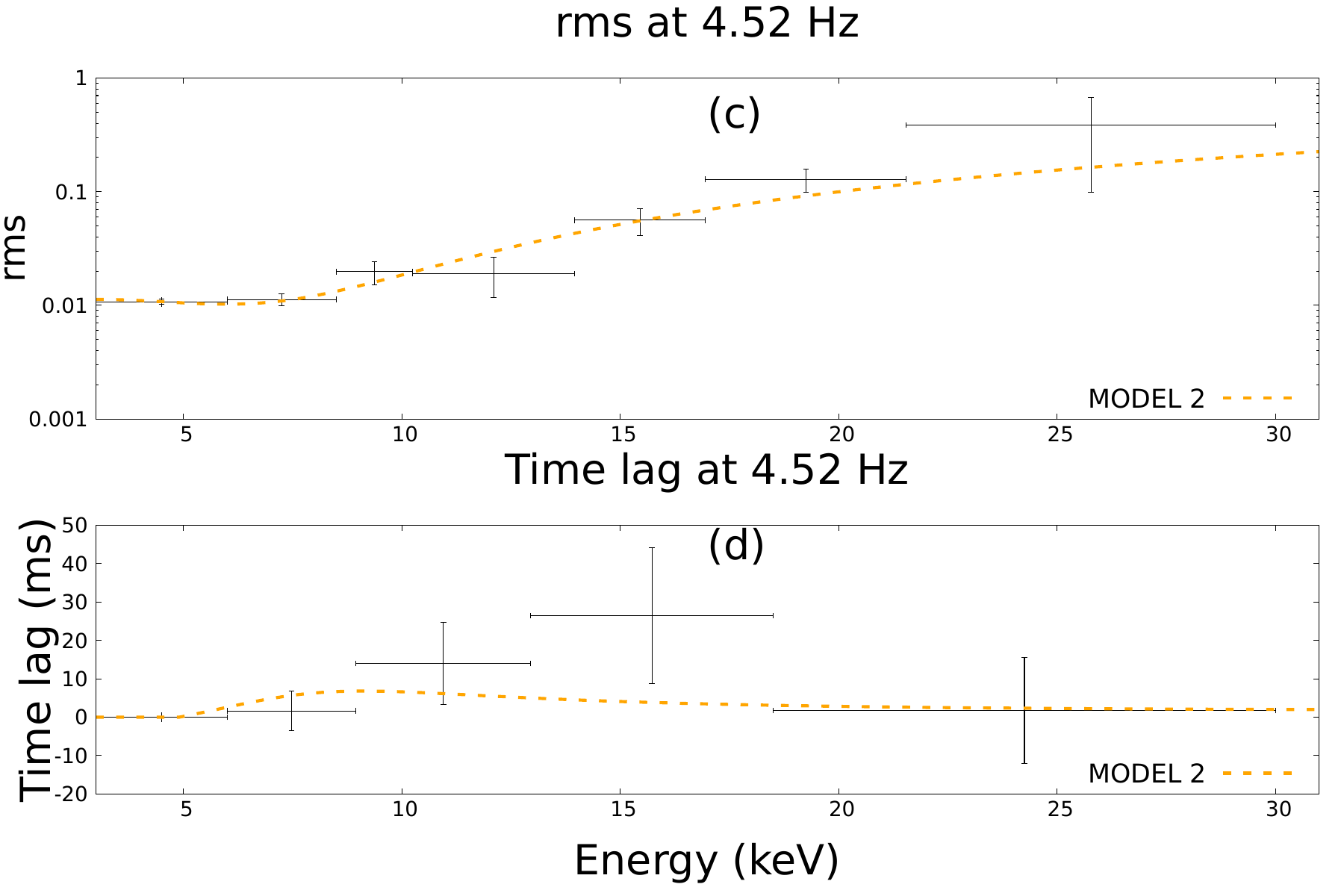}
		\includegraphics[width=0.45\linewidth]{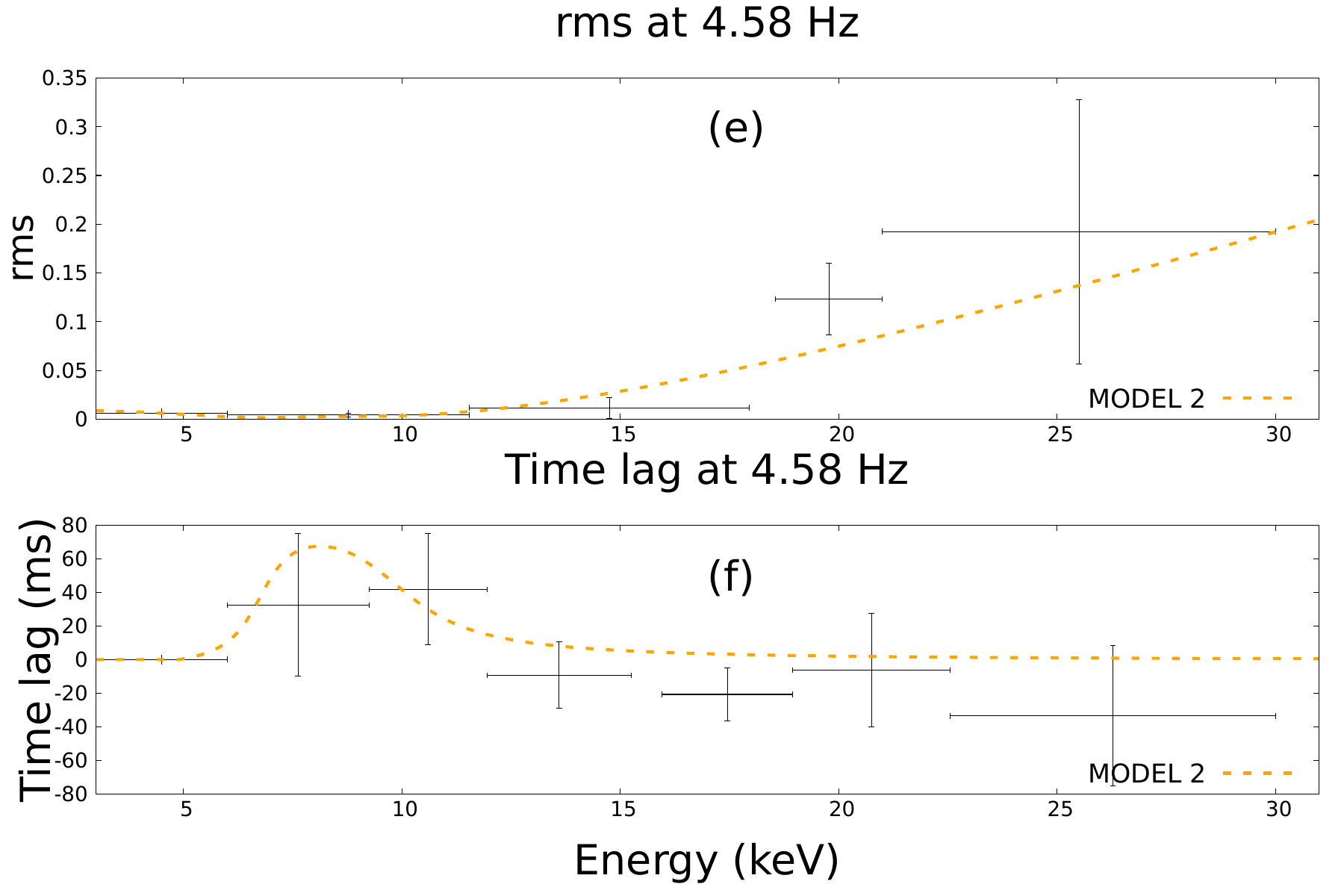}
				\includegraphics[width=0.45\linewidth]{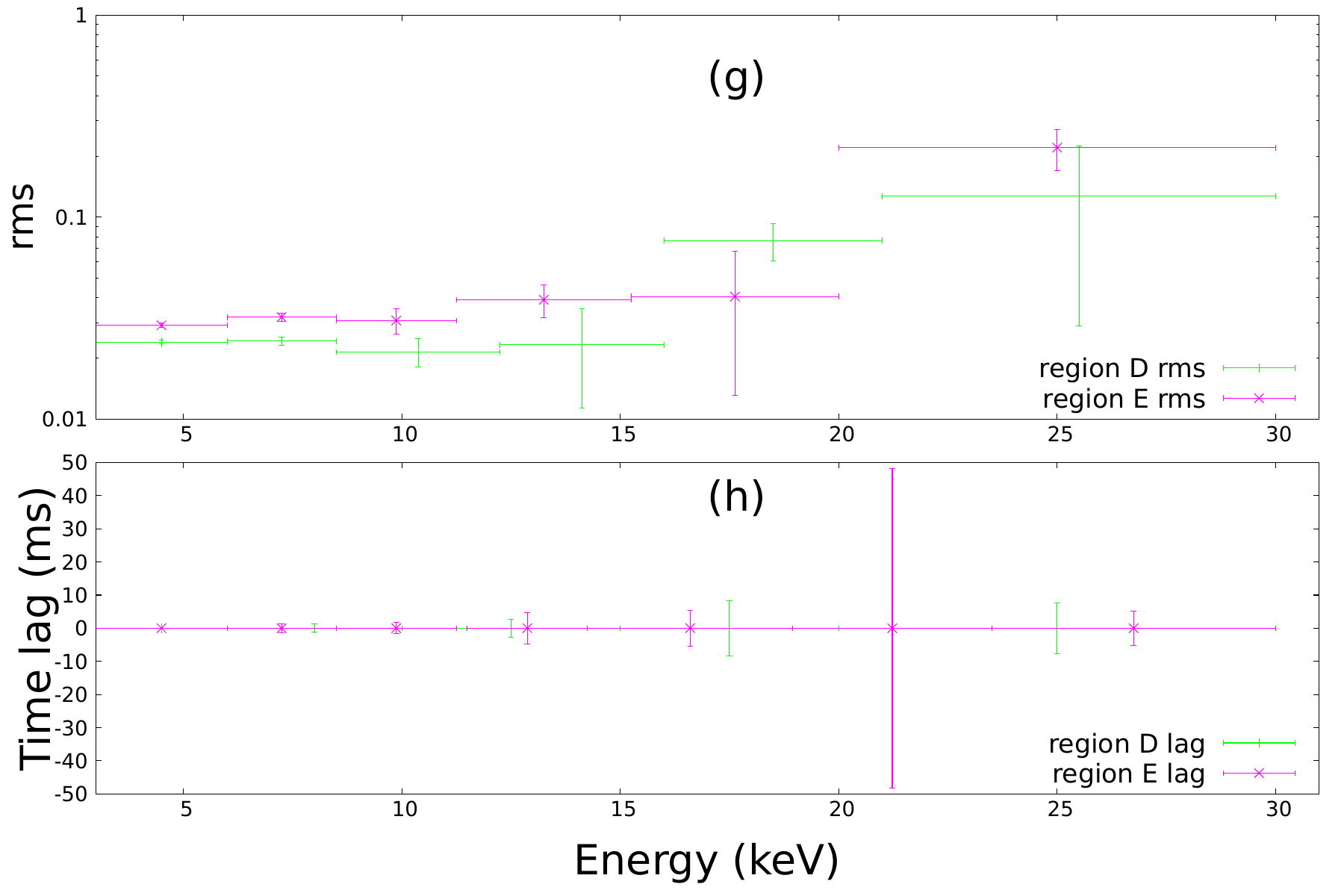}
		\caption{ The fractional rms and time lag of Region A ((a) and (b)), Region B ((c) and (d)), and Region C ((e) and (f)) are fitted with MODEL 2, using a reference energy ($E_{\rm ref}$) of 4.5 keV. The fractional rms and lag for Regions D and E are shown in panels (g) and (h), respectively.}
		\label{RMS_LAG}
	\end{figure}

		The phase lag terms such as $\phi_{\tau}$ and $\phi_{\rm \dot{H}}$ were not needed in the fitting, suggesting that the variations occur simultaneously. Hence, MODEL 2 consists of three free parameters. When fitting MODEL 2 to the rms and lag of Regions B and C, we obtained the best fit with reduced $\chi^2$ values of approximately 1.14 (9.12/8) and 1.22 (9.78/8), respectively. Interestingly, MODEL 2 provided a good fit for both the rms and lag spectra of all three regions without requiring any phase lag or delay terms. We believe that this may be due to the relatively insignificant variations in the lags. Therefore, we did not attempt to vary further parameters due to the risk of overfitting. Figure \ref{RMS_LAG} shows the model fitting on the rms and lag of different regions, which were generated at the QPO frequency. In MODEL 2, the parameters exhibit variation, yet the lag terms are unnecessary to explain the energy-dependent rms and lag. This outcome likely stems from the minimal variation in lag, a phenomenon also observed in the NBO and discussed by Pahari et al. \cite{11x}. Their study attributes the insignificance of lags to poor statistics, a limitation that likely obscured significant lag variations. In the present work, with improved statistics, we anticipate that the observed lags in all regions might exhibit more pronounced variations. It is clear that there is only a marginal variation in the lag across all regions. Due to this marginal variation, it is challenging to determine the true nature of the lags in these regions. Additionally, the model fitting of the lag does not exhibit the proper trend of lags. In Regions D and E, we did not employ the propagation model due to the zero variation in the lag, as shown in Figure \ref{RMS_LAG} ((g) and (h)).

The best-fit model parameters are listed in Table \ref{S1_S2}. The fitting of rms and time lag using MODEL 2 indicates that the corona probably plays a role in the generation of QPOs in this system, together with fluctuations occurring on the NS surface or within the boundary layer. Our findings suggest that variability in the system arises from changes in the blackbody temperature associated with the NS surface or boundary layer as well as from alterations in coronal heating and optical depth. Based on the MODEL 2 parameters for Region A, we believe that there were simultaneous variations in blackbody temperature, coronal heating rate, and optical depth. A similar interpretation can be drawn from the fitting results for Regions B and C. The non-involvement of the phase term in the model may be due to the marginal variations in the lag across all regions.

	{\section{Results and Discussion}{\label{DC}}}

	We investigated the low-frequency oscillations in GX 13+1, which we detected using \astrosat{} observations. We performed flux-resolved spectroscopy, following the same method as Giridharan et al.~\cite{3k}. We have reproduced the spectral parameters and used them to model the rms and lag extracted for different regions. 
	
	The variability seen in the XBs are attributed to the propagation of stochastic fluctuations generated at various radial positions within the accretion disk. These perturbations can move inward through viscous diffusion processes~\cite{6,16h}, or outward, resulting in soft lags that can be described using the Green’s function formalism~\cite{5}. Hard X-ray emission primarily originates from the innermost regions of the system, likely due to elevated temperatures in the inner disk or the presence of a hot corona, leading to time lags that occur on viscous timescales~\cite{10}. The direction of fluctuation propagation, whether inward or outward, depends on the physical mechanism involved and is often associated with the sound-crossing timescale~\cite{7}. Maqbool et al.~\cite{16h} proposed a model in which variations at the inner boundary of a truncated disk propagate toward the hot inner flow after a certain delay, as demonstrated in the BHXB Cygnus X-1. This framework was used to interpret both the rms variability and the time lags observed in the BBN. With a similar interpretation, Garg et al.~\cite{12-a,16g} proposed a more generic approach to explain the time lag and rms of QPOs and BBNs observed in XB systems.
	
	In our work, we focused on the lower frequency oscillations/BBN upon revisiting this observation. Interestingly, we detected the presence of low-frequency QPOs. In Regions A, B, and C, we detected QPOs that show similar characteristics to the NBO usually seen in Z-track sources. Similar QPO frequencies and features have been reported in other Z-track sources, such as Cyg X-2 ~\cite{11k,11j} and GX 340+0 ~\cite{11x}.
	
	The physical process of the origin of NBOs is highly ambiguous. These QPOs are also believed to be linked to variations in the optical depth of a radially inflowing spherical region, potentially driven by radiation pressure feedback operating near the Eddington luminosity limit~\cite{11g,11h}. Furthermore, M. A. Alpar et al.~\cite{11i} proposed that NBOs may arise from acoustic oscillations within a geometrically thick accretion disk. Investigating energy-dependent timing variations, such as rms and lag of X-rays, can provide insight into the connection between spectral components and offer information about the physical origin of these variabilities.
	
With that goal, we modeled the rms and lag of the detected QPOs using the propagation model proposed by Garg et al.~\cite{12-a,16g}. The rms and lag were modeled with MODEL 2, which consists of variations in blackbody temperature, heating rate, and optical depth. The rms showed an increasing trend with energy, while the variation in lag showed an interesting behavior. Taking into account the statistical uncertainties in the coherence and lag values for Regions A and C, we have modeled the time lags for all three regions. This lag modeling is meaningful within the context of our analysis, as we simultaneously model both the energy-dependent fractional rms and the time lag. Notably, MODEL 2 was able to fit the rms and lag data for all three regions without requiring any additional phase lag or delay terms. We believe that this is likely due to the negligible variations in the lag values. As shown in Figure \ref{RMS_LAG}, the lags across all regions exhibit minimal variation. Initially, the lag values increase toward positive values; however, beyond a certain energy band, they transition to soft lags—this behavior is particularly evident in Regions A (Figure \ref{RMS_LAG}(b)) and C (Figure \ref{RMS_LAG}(f)). In contrast, the lags in Region B show more pronounced variation across specific energy bands. A similar trend in the evolution of NBO time lags was previously reported by Sudha et al.~\cite{11k} for the Z-source Cyg X-2.
	
The modeling of rms and lag of the QPOs provides information on the involvement of the corona and the boundary layer or the NS surface. The MODEL 2 used to fit the rms and lag can be understood as the variation in the boundary layer or NS surface temperature and the variation in the coronal heating rate, along with the variation in optical depth, occurring simultaneously and causing variability in this system. The variation in optical depth in our modeling can be implied as the main cause behind the NBOs in this system. These variations may cause oscillations in the optical depth of the corona. To determine exactly what causes the production of NBOs, more NBO-detected sources should be studied, considering both disk emission and boundary layer or NS surface emission.

{\section{Conclusions}{\label{Con}}}
	We analyzed the \astrosat{} observation of GX 13+1 to understand the low-frequency variability with the propagative model. The main findings from our work are:
\begin{enumerate}
	\item We detected the presence of low-frequency oscillations in Regions A, B, and C. These oscillations were detected around at a frequency of $\sim$ 5 Hz in these three regions.
	\item Upon investigating the characteristics of these oscillations, we confirm that they are QPOs, probably NBOs, which are generally detected in the normal branch of Z-track sources, such as GX 340+0 ~\cite{11x} and GX 5-1 ~\cite{11l}. This may be the first detection of NBO-like QPOs in GX 13+1. Coordinated long-duration observations of GX 13+1 across multiple epochs by missions such as {\it AstroSat} and {\it NICER} would be valuable for further investigating the presence of these QPOs and for gaining a better understanding of the HID.
	\item The extracted rms showed an increasing trend with energy, while the lags exhibited marginal variations. The lags were initially positive and evolved towards negative; in Region A, they evolved towards positive again. Similar trends have also been reported by Sudha et al.~\cite{11k} for NBOs.
	\item By investigating the energy-dependent properties of the rms and lag using a propagative model, we found that the variations behind the production of these QPOs are due to variations in blackbody temperature, coronal heating rate, and optical depth.

\end{enumerate}

More work needs to be carried out to properly understand the origin of low-frequency oscillations in NSXBs. Recently, in a study on the LM-NSXB 4U 1608-52~\cite{9}, the BBN variability was studied, and its origin was explained using the Garg et al.~\cite{12-a} model formalism, which attributed it to perturbations in the heating rate and inner-disk temperature. The technique explained in Garg et al.~\cite{12-a} needs to be applied to more NSXB systems, combining models with \texttt{bbodyrad}, to understand the origin of different oscillations in these systems. Z-track sources are one of the facilitating sources, and the origin of the variability observed in these systems is still a topic of research. Therefore, incorporating the propagative model, such as Garg et al.~\cite{12-a}, in sources such as GX 340+0, Cyg X-2, GX 17+2, and GX 5-1 can provide new insights into these features.

	{\bf Acknowledgements:}
	We acknowledge the High Energy Astrophysics Science Archive Research Centre (HEASARC) for the software used in this work. Additionally, we have employed data from the \astrosat{} (LAXPC/SXT) mission. We thank the LAXPC Payload Operation Center (POC) and the SXT POC at TIFR, Mumbai, for providing archival data and necessary software tools. We acknowledge the MAXI RIKEN team for the MAXI data used in this research work. A.P. sincerely thanks Dr. Akash Garg, IUCAA, Pune, for insightful discussions on the model implications. A.P. acknowledges financial support through a fellowship from Tezpur University, Assam. S.B. acknowledges financial support from the DST-INSPIRE (Grant No.: DST/INSPIRE Fellowship/[IF220164]) fellowship by the Government of India. B.S. would like to acknowledge the Visiting Associateship Programme of IUCAA, Pune. We express our gratitude to the anonymous referees for their constructive comments and recommendations, which have improved the quality of the manuscript.

\end{document}